\documentclass{article}
\usepackage{fix-cm}

\usepackage{PRIMEarxiv}

\usepackage[utf8]{inputenc} 
\usepackage{newtxtext}
\usepackage[T1]{fontenc}    
\usepackage{hyperref}
\hypersetup{
  colorlinks=true,
  linkcolor=red,
  filecolor=black,  
  citecolor=cyan,
  urlcolor=blue,
  pdftitle={Overleaf Example},
  pdfpagemode=FullScreen,
  }
\usepackage{url}            
\usepackage{booktabs}       
\usepackage{amsfonts}       
\usepackage{nicefrac}       
\usepackage{microtype}      
\usepackage{lipsum}
\usepackage{fancyhdr}       
\usepackage{graphicx}       
\graphicspath{{media/}}     
\usepackage{multirow}
\usepackage{float}
\usepackage{amsmath} 
\usepackage{xcolor}

\usepackage{algorithm}
\usepackage[linesnumbered,ruled,algo2e]{algorithm2e}
\usepackage{caption}
\renewcommand{\thealgocf}{}   

\title{A Faster Algorithm for Maximum Weight Matching on Unrestricted Bipartite Graphs}

\author{
 Shawxing Kwok \\
 Independent Researcher\\ 
 Anhui, China \\ 
\texttt{shawxingkwok@126.com}
}

\begin{document}
\fontsize{11pt}{12pt}\selectfont
\maketitle

\begin{abstract}
Given a weighted bipartite graph $G = (L, R, E, w)$, the maximum weight matching (MWM) problem seeks to find a matching $M \subseteq E$ that maximizes the total weight $\sum_{e \in M} w(e)$.  

This paper presents a novel algorithm with a time complexity of $O(\min(X^3 + E, XE + X^2\log X))$, where $X = \min(|L|, |R|)$. Unlike many existing algorithms, our approach supports real-valued weights without additional constraints. Under this condition, our result improves upon the previous best-known bound of $O(VE + V^2\log V)$, or more strictly $O(XE + XV\log V)$, where $V = L \cup R$.  

The suggested implementation code is simplified and publicly available at \url{https://github.com/ShawxingKwok/Kwok-algorithm}, with the average-case time complexity of $O(E^{1.4} + LR)$ estimated from experimental results on random graphs.
\end{abstract}
 
\section{Introduction} \label{sec:Introduction}
In many real-world scenarios, entities must be optimally paired based on weighted preferences. A typical example is the job assignment problem, where a set of workers $L$ must be assigned to a set of tasks $R$, with each worker-task pair associated with a certain compatibility score or cost. The goal is to maximize the overall efficiency of the assignment while ensuring that no worker is assigned to multiple tasks.

This kind of problem can be abstracted as the maximum weight matching (MWM) problem, a fundamental challenge in combinatorial optimization with applications in various fields, such as deep learning and autonomous driving. Formally, it is defined on a weighted bipartite graph $G = (L, R, E, w)$, where $L$ and $R$ are disjoint vertex sets, $E$ is the set of edges connecting vertices in $L$ to vertices in $R$, and $w: E \to \mathbb{R}$ assigns a weight to each edge. A matching $M \subseteq E$ is a subset of edges such that no two edges share a common vertex. The objective is to find a matching $M$ that maximizes $\sum_{e \in M} w(e)$. Additionally, in a perfect matching, each vertex is matched, implying $|L| = |R|$ and leading to a subproblem known as the maximum weight perfect matching (MWPM) problem.

Typically, it is assumed that $O(L) = O(R) = O(V)$. However, the case where $O(L) \neq O(R)$ is common and deserves further consideration. For ease of analysis, we ensure $|L| \leq |R|$ by designating the side with fewer vertices as $L$ and the side with more vertices as $R$ during graph construction. Meanwhile, the condition $|E| \geq |R|$ can also be satisfied by removing vertices that have no incident edges. Thus, before applying the proposed algorithm in practice, the graph may need to be reconstructed. However, the time required for this preprocessing step is negligible. \textbf{In addition, $X$ in the abstract is replaced by $|L|$ subsequently.}

The MWM and MWPM problem has been extensively studied and refined, with contributions from many researchers in the fields of mathematics, computer science, and operations research. Over the years, various algorithms have been introduced, as summarized in Table~\ref{tab:algorithms} with uniform notes as follows:
\begin{itemize}
    \item $N$ denotes the maximum edge weight.
    \item Algorithms designed for MWPM inherently solve MWM with the same time complexity by disregarding negatively weighted edges and introducing virtual vertices and edges with zero weight to ensure $|L| = |R|$ and $|E| = |L||R|$.
    \item MWM algorithms can address MWPM by adding a constant to all edge weights to ensure non-negative weights and then each vertex is matched, while the factor $N$ in the running time transforms into $|V|N$.
    \item Some algorithms tackle more generalized problems, with MWM as a subproblem.
\end{itemize}

\renewcommand{\arraystretch}{1.5} 
\begin{table}[H]
  \caption{Historical results on MWPM and MWM.}
  \label{tab:algorithms}
  \centering
  \resizebox{\textwidth}{!}{%
  \begin{tabular}{ccccc}
    \toprule
    Year & Author & Problem & Upper Time Bound & Notes \\
    \midrule
    From 1955 & Kuhn, Munkres et al \cite{Kuhn1955} \cite{Munkres1957} & MWM & $L^2R$ & Hungarian algorithm with line covering \\
    \hline
    1971 & Tomizawa \cite{tomizawa1971some}&  \multirow{2}{*}{\centering MWM } & \multirow{2}{*}{\centering $L \cdot SP^+$ } & \multirow{2}{*}{\shortstack{$SP^+$\ =\ time\ for\ SSSP\ on\ a\ \\ non-negatively\ weighted\ graph}} \\
    1972 & Edmonds \& Karp  \cite{edmonds1972theoretical} & & & \\
    \hline
    1976 & Bondy \& Murty \cite{Bondy1976} \cite{Cormen2022} & MWPM & $V^3$ & Hungarian algorithm without line covering \\
    \hline
    1985 & Gabow \cite{gabow1985scaling} & MWM & $EV^{3/4} \log N$ & Integer weights \\
    \hline
    1986 & Galil, Micali \& Gabow \cite{galil1986ev} & MWM & $EV\log V$ & \\
    \hline
    1987 & Fredman \& Tarjan \cite{fredman1987fibonacci} & MWM & $VE + V^2\log V$ & More precisely $LE + LV\log V$ \\
    \hline
    1989 & Gabow \& Tarjan \cite{gabow1989faster} & MWM & $\sqrt{V}E \log(VN)$ & Integer weights \\
    \hline
    2002 & Thorup \cite{thorup2002} & MWM & $VE$ & Integer weights, randomized \\
    \hline
    2003 & Thorup \cite{thorup2003} & MWM & $VE + V^2 \log \log V$ & Integer weights \\
    \hline
    2006 & Sankowski\cite{sankowski2006weighted} & MWM & $NV^\omega$ & \shortstack{Integer weights, randomized, \\ $\omega =$ matrix mult. exponent} \\
    \hline
    2012 & Ran Duan \& Hsin-Hao Su \cite{DuanRan2012} & MWM & $\sqrt{V}E \log N$ & Integer weights \\
    \hline
    2019 & M. Asathulla et al \cite{asathulla2019faster} & MWPM & $V^{4/3}\log\ VN$ & Integer weights \\
    \hline
    2022 & Li Chen et al \cite{chen2022maximum} & MWM & $E^{1 + o(1)}$ & Polynomial integer weights  \\
    \hline
         & New result & MWM & $\min(L^3+E,\ LE + L^2\log L)$ & \\
    \bottomrule
  \end{tabular}%
  }
\end{table}


\subsection{Solution with Min-Cost Max-Flow} \label{sec:MCMF}
It was known that the assignment problem can be reduced to $|L|$ shortest path computations on arbitrarily weighted graphs with negating all the original edge weights. Ford and Fulkerson \cite{ford1962flows}, Hoffman and Markowitz \cite{hoffman1963note}, and Desler and Hakimi \cite{desler1969graph} provided different reductions in the 1960s.  

In 1971 and 1972, Tomizawa \cite{tomizawa1971some} and Edmonds and Karp \cite{edmonds1972theoretical} further analyzed the min-cost max-flow problem. The approach is first to determine the initial potential using an SSSP method on an arbitrarily weighted graph, and then execute an SSSP method multiple times on a non-negatively weighted graph. Dijkstra’s algorithm is commonly the best choice for the second SSSP method. Consequently, for the MWM problem, the transformed graph for the min-cost max-flow problem is initially acyclic, allowing us to determine the initial potential in $\Theta(E)$ time. Then, we can obtain the MWM after executing Dijkstra’s algorithm $|L|$ times, since each execution increases the matching by one pair. 

In 1987, Fredman and Tarjan introduced the Fibonacci heap \cite{fredman1987fibonacci}, enabling Dijkstra’s algorithm to run in $O(E + V \log V)$ time. Since then, this has remained the theoretically fastest known implementation of Dijkstra’s algorithm for graphs with real-valued edge weights. As a result, the MWM problem can be solved in $O(L(E + V \log V)) = O(LE + LV \log V)$ time. Note that since Dijkstra’s algorithm is fully executed in each iteration to update all the potentials, the average-case time complexity is expected to be $\Theta(LE + LV \log V)$ with high probability.

\subsection{Hungarian Algorithm}
Among algorithms listed in Table~\ref{tab:algorithms}, the Hungarian algorithm stands out as the most widely adopted for the superior practical performance, especially for dense graphs. It was initially developed by Harold Kuhn\cite{Kuhn1955} in 1955 for the MWPM problem. Kuhn named the algorithm "Hungarian algorithm" in honor of the foundational work by Hungarian mathematicians Dénes Kőnig and Jenő Egerváry, who laid the theoretical groundwork for bipartite matching in the early 20th century. Subsequent optimizations, including those by Munkres \cite{Munkres1957} and others over different periods, reduced its complexity to $O(V^3)$. This algorithm is also known as the Kuhn-Munkres (KM) algorithm, named in honor of Kuhn and Munkres. A key aspect of the principle involves covering certain rows and columns with lines; therefore, we refer to it as the line-covering method in this paper. As the caption of table 1 said, the line-covering method can run in $O(V^3)$ time for the MWM problem through graph augmentation. However, it has been later optimized, eliminating the need to add vertices and achieving a time complexity of $O(L^2R)$. In 1976, Bondy and Murty\cite{Bondy1976}  \cite{Cormen2022} proposed a non-line-covering variant, which is commonly regarded as having an $O(V^3)$ time complexity. Both the line-covering method and its non-line-covering variant continue to be widely applied.

This paper focuses on the non-line-covering variant of the Hungarian algorithm for the MWM problem, achieving the following results:  
\begin{enumerate}
    \item When virtual vertices are added as $L'$ to ensure $|L| + |L'| = |R|$, this paper proves that the time complexity is more precisely $O(LR^2)$, less than $O(V^3)$.  
    \item It further proves that adding vertices is unnecessary, reducing the time complexity from $O(LR^2)$ to $O(L^2R)$.  
    \item The variant is extended to eliminate the need for edge expansions, significantly improving computational efficiency. The proposed algorithm achieves a worst-case time complexity of $O(\min(L^3 +E,\ LE + L^2\log L))$ and exhibits superior average-case performance in random instances, particularly for sparse graphs.  
\end{enumerate}

Next begins by explaining the principles underlying the original non-line-covering variant of the Hungarian algorithm, followed by optimizations, extensions and concluding with detailed experimental results.

\section{Definitions and Preliminaries}

\subsection{Complete Weighted Bipartite Graph Example}

\begin{figure}[H]
    \centering
    \includegraphics[width=.8\textwidth]{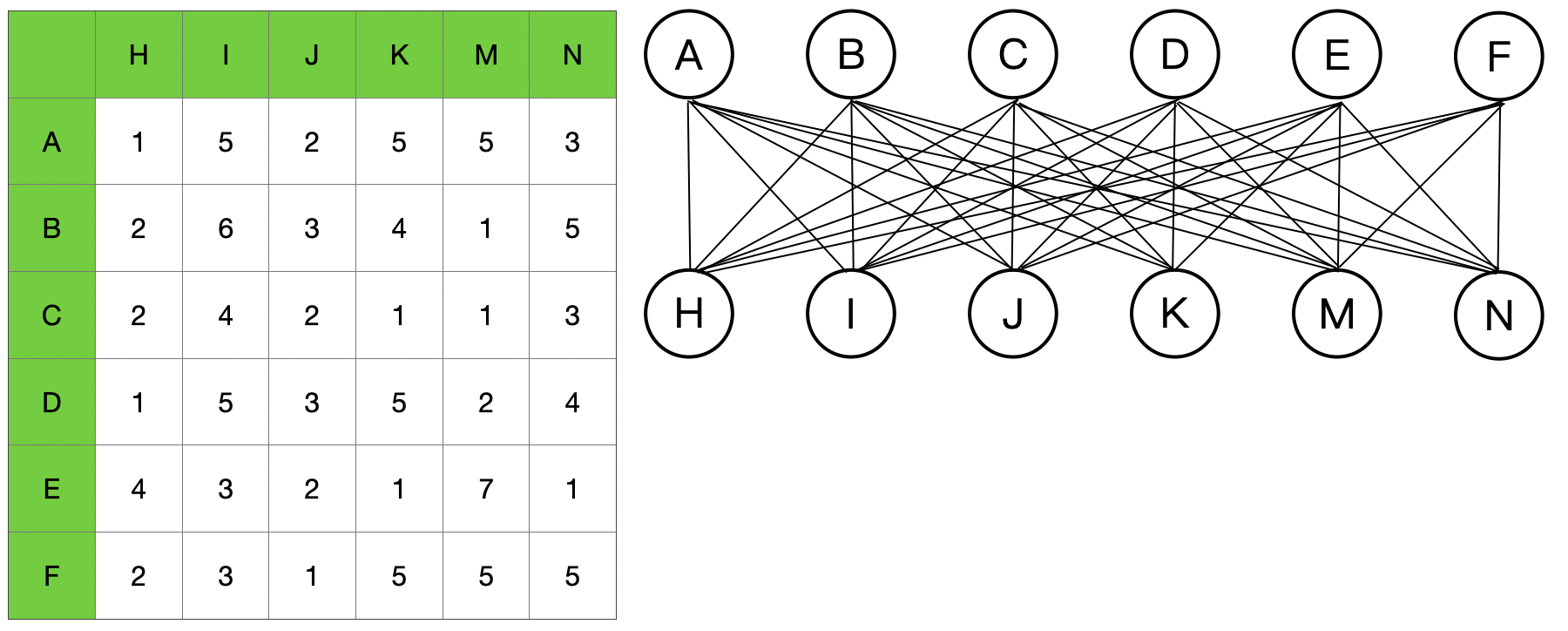}
    \caption{Adjacency matrix and corresponding bipartite graph $G = (L, R, E, w)$.}
    \label{fig:background}
\end{figure}

\begin{itemize}
    \item To ensure a visually comfortable presentation, this paper adopts the convention that lowercase letters in the text correspond to uppercase letters in the diagrams.
    \item $L = \{a, b, c, d, e, f\}$
    \item $R = \{h, i, j, k, m, n\}$
    \item $\forall l_1, l_2 \in L, (l_1, l_2) \notin E;\ \forall r_1, r_2 \in R, (r_1, r_2) \notin E$
    \item $\forall l \in L, r \in R, (l, r) \in E$
    \item The weights $w$ are stored in the adjacency matrix. 
\end{itemize}

\subsection{M-Augmenting Path $p^{\uparrow}$}

\begin{figure}[!http]
    \centering
    \includegraphics[width=.9\textwidth]{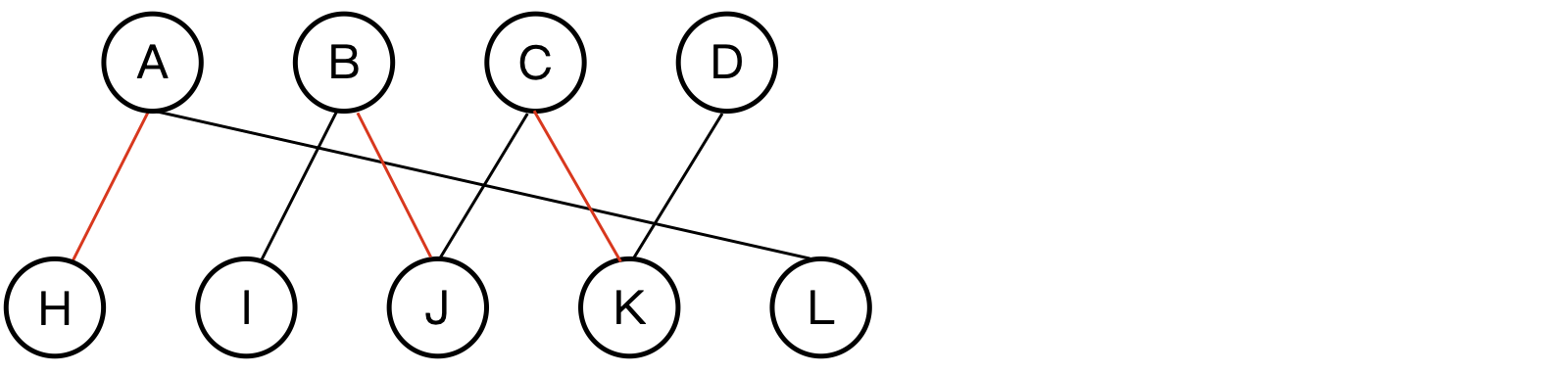}
    \caption{Initial matching $M$ in $G$.}
    \label{fig:aug1}
\end{figure}

In figure~\ref{fig:aug1}, let $G = (L, R, E)$, $L = \{a, b, c, d\}, R = \{h, i, j, k, l\}$, $E$ consists of both black and red edges. The set of red edges represents the matching $M$.

Clearly, by replacing $(b, j)$ and $(c, k)$ in $M$ with $(b, i), (c, j)$, and $(d, k)$, we can obtain a larger available matching. These five edges form the path $\langle i, b, j, c, k, d \rangle$, which satisfies the following properties:

\begin{enumerate}
    \item The first and last vertices are unmatched.
    \item Black and red edges alternate along the path.
    \item The first and last edges are black.
\end{enumerate}

\begin{figure}[!http]
    \centering
    \includegraphics[width=.9\textwidth]{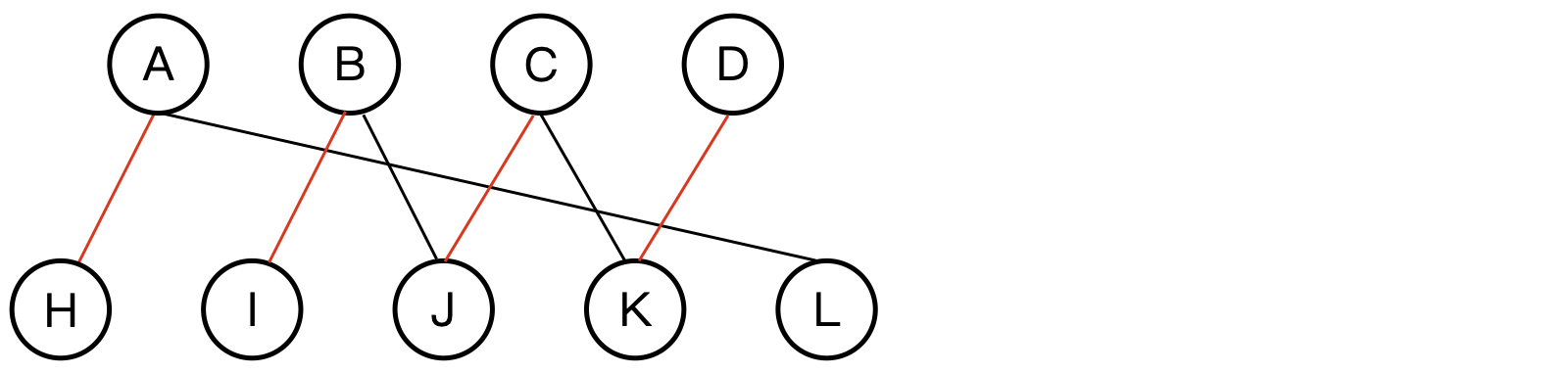}
    \caption{Graph after applying $p^{\uparrow}$.}
    \label{fig:aug2}
\end{figure}

This path is referred to as an \textbf{$M$-augmenting path}. Once the path is applied, the graph changes as illustrated in Figure~\ref{fig:aug2}.

For convenience, this paper designates such a path as $p^{\uparrow}$. It is important to note that $p^{\uparrow}$ is not unique within a bipartite graph. In the more complex example below, another $M$-augmenting path, $p^{\uparrow} = \langle i, b, j, d, l, c \rangle$, can be similarly identified.

\begin{figure}[H]
    \centering
    \includegraphics[width=.9\textwidth]{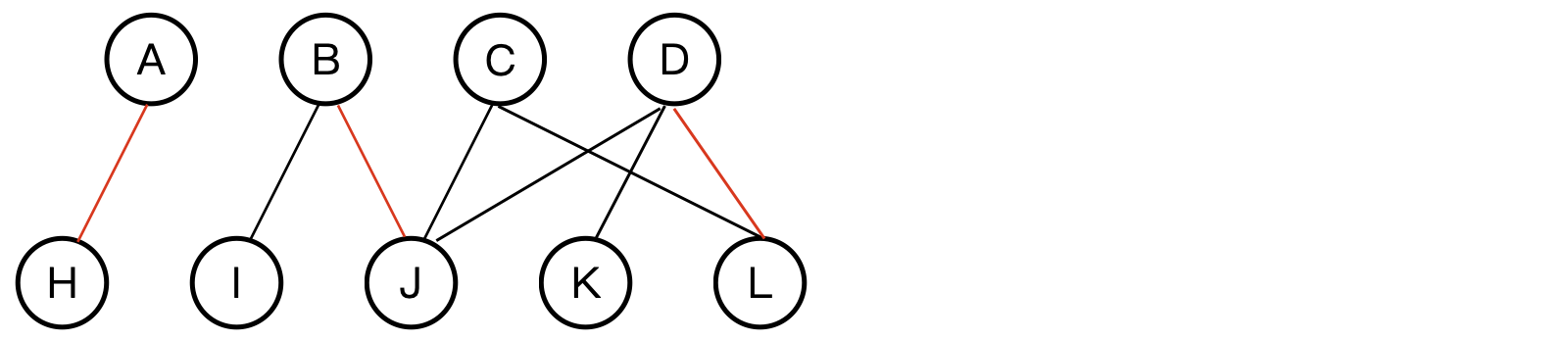}
    \caption{Example of a complex $p^{\uparrow}$.}
    \label{fig:aug3}
\end{figure}

In addition, $p^\uparrow$ may contain only a black edge.

\subsection{Equality Subgraph} \label{sec:EqualitySubGraph}

\begin{figure}[H]
    \centering
    \includegraphics[width=.9\textwidth]{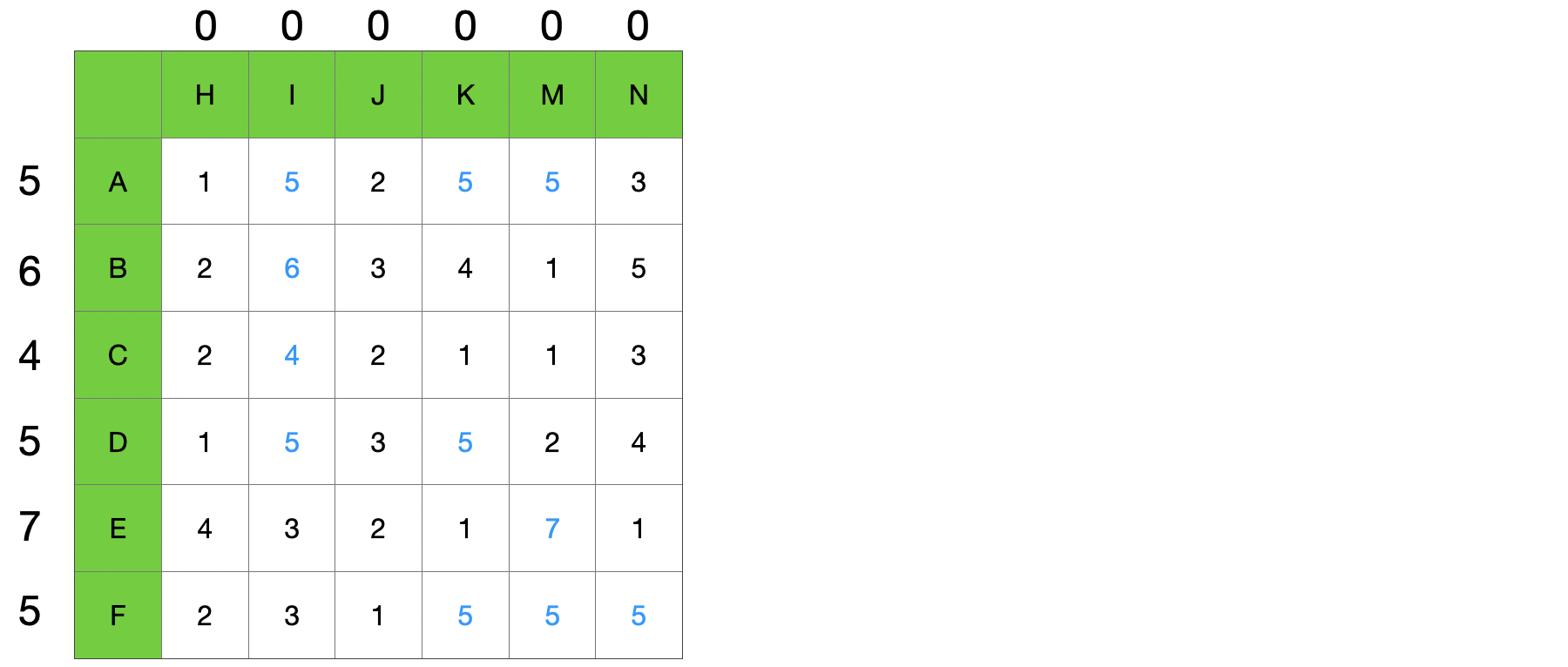}
    \caption{Initial equality subgraph with vertex labeling.}
    \label{fig:equality}
\end{figure}

Figure~\ref{fig:equality} illustrates an initial vertex labeling $h$ with the adjacency matrix defined as follows:
\begin{itemize}
    \item $\forall l \in L,\ l.h = \max\{w(l, r): r \in R\}$
    \item $\forall r \in R,\ r.h = 0$
\end{itemize}
For instance, $a \in L$ and $a.h = 5$, $j \in R$ and $j.h = 0$.

\begin{figure}[H]
    \centering
    \includegraphics[width=.9\textwidth]{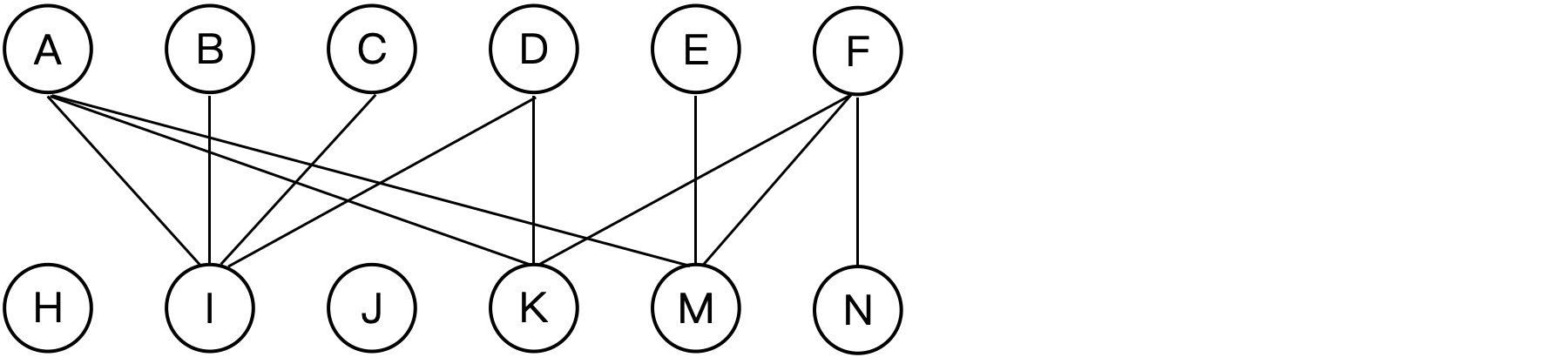}
    \caption{Equality subgraph $G_h$ showing feasible edges.}
    \label{fig:equalityGraph}
\end{figure}

Let $E_h = \{(l, r): (l, r) \in E \ \text{and} \ w(l, r) = l.h + r.h\}$. Figure~\ref{fig:equalityGraph} shows the initial equality subgraph $G_h = (L, R, E_h, w)$, in which each edge corresponds to a blue number in the adjacency matrix in Figure~\ref{fig:equality}.

Subsequently, the equality subgraph varies with $l.h$ and $r.h$, but the condition $l.h + r.h \geq w(l, r)$ must always hold. We refer to $h$ as a \textbf{feasible vertex labeling}.

\textbf{Theorem 2.3.1}. \label{theorem231} After iteratively adjusting $h$ and enlarging the matching, if a perfect matching $M^\star$ is found in $G_h$, then $M^\star$ is also an MWPM in the original graph $G$.

\textit{Proof}. Let $M$ be an arbitrary perfect matching in $G$. Since every vertex is matched in a perfect matching, we have $|M| = |L| = |R|$. Given the inequality $w(l, r) \leq l.h + r.h$, we can conclude that:

\begin{equation}
    w(M) = \sum\limits_{(l,r)\in M}w(l,r) \leq \sum\limits_{(l,r)\in M}(l.h + r.h) = \sum\limits_{l\in L}l.h + \sum\limits_{r\in R}r.h
    \label{eq:inequality1}
\end{equation}

Now assume $M^\star$ represents a perfect matching within the subgraph $G_h$. Since $w(l, r) = l.h + r.h$ for each $(l, r) \in E_h$ and $|M^\star| = |L| = |R|$, we have:

\begin{equation}
    w(M^\star) = \sum\limits_{(l,r)\in M^\star}w(l,r) = \sum\limits_{(l,r)\in M^\star}(l.h + r.h) = \sum\limits_{l\in L}l.h + \sum\limits_{r\in R}r.h
    \label{eq:inequality2}
\end{equation}

Therefore:
\begin{equation}
    w(M) \leq \sum\limits_{l\in L}l.h + \sum\limits_{r\in R}r.h = w(M^\star)
    \label{eq:inequality3}
\end{equation}

This demonstrates that $M^\star$ is also an MWPM in $G$.

\section{Non-Line-Covering Variant of the Hungarian Algorithm} \label{sec:varaint}

\subsection{Procedure} \label{sec:procedure}

\begin{figure}[!http]
    \centering
    \includegraphics[width=.9\textwidth]{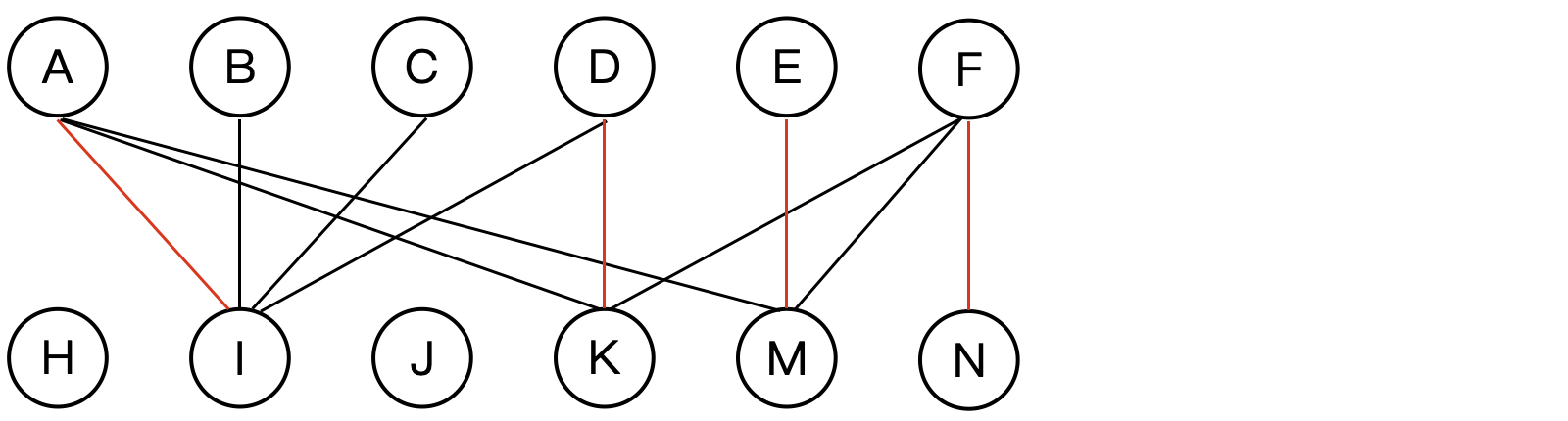}
    \caption{Initial selection of matchings in $G_h$.}
    \label{fig:initial-matching}
\end{figure}

Figure~\ref{fig:initial-matching} illustrates an initial greedy selection of matching from $G_h$, which is unnecessary but significantly accelerates the process. 

\begin{figure}[!http]
    \centering
    \includegraphics[width=.9\textwidth]{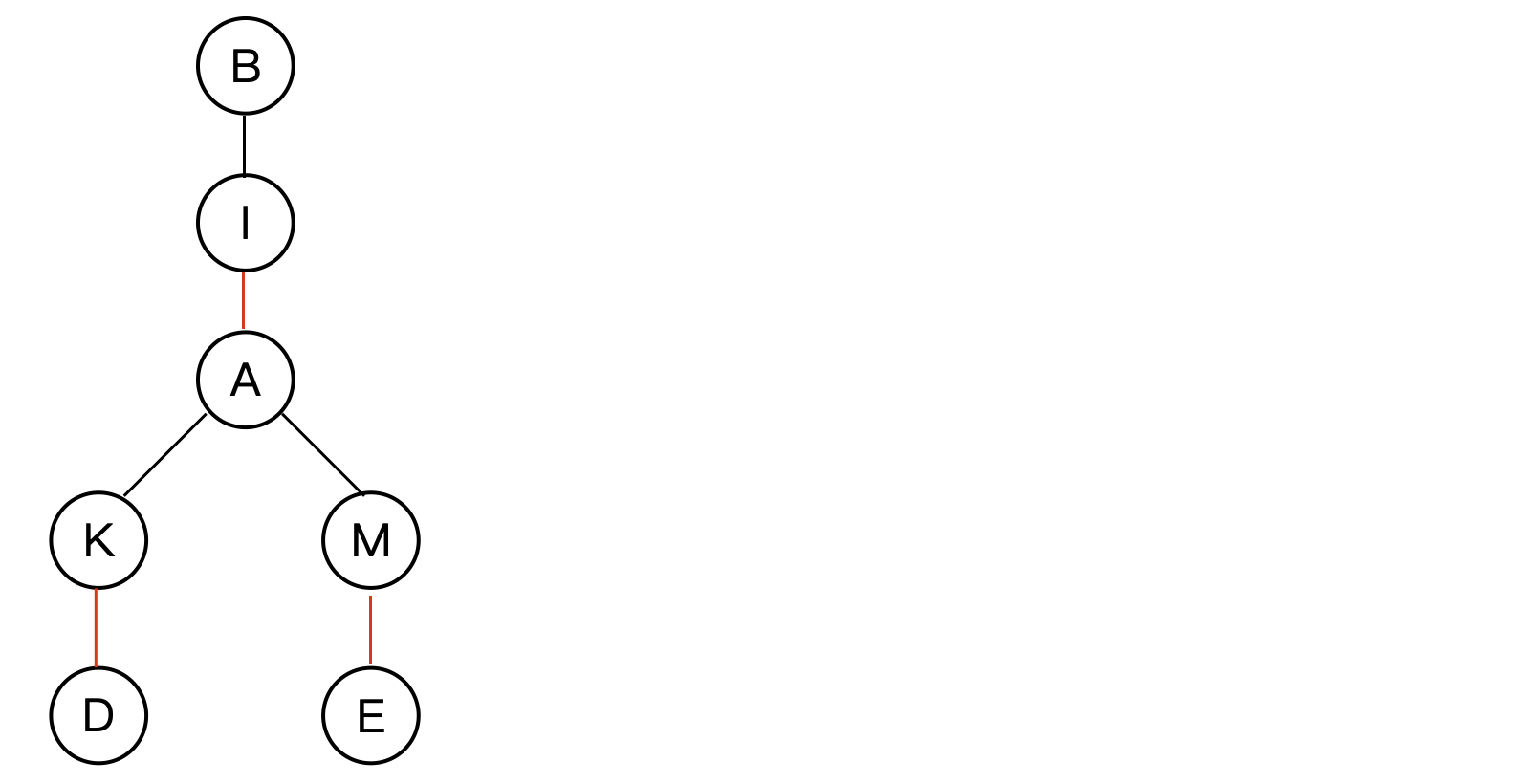}
    \caption{Breadth-first tree starting from $b$.}
    \label{fig:bfs-tree}
\end{figure}

Next, for each $l \in L$, if $l$ is unmatched, a breadth-first search (BFS) is performed with $l$ as the source to search for $p^{\uparrow}$ with the emphasized principle of alternating black and red edges. For instance, when $b$ is enumerated, the breadth-first tree shown in Figure~\ref{fig:bfs-tree} emerges. In this scenario, no $p^{\uparrow}$ is found. To increase $|M|$, it's necessary to extend $E_h$.

Let this limited breadth-first tree be $T$, where the vertex sets in this example on both sides are $T.L = \{a,b,d,e\}$ and $T.R = \{i,k,m\}$, with the edge set given by $T.E = \{b \rightarrow i, i \rightarrow a, a \rightarrow k, a \rightarrow m, k \rightarrow d, m \rightarrow e\}$. To obtain $p^{\uparrow}$, additional edges should be introduced into $E_h$, with the selection range defined as $\{ l \rightarrow r : l \in T.L, \ r \in R - T.R \}$. Specific restrictions will be detailed later. BFS continues, and $T$ can then expand by incorporating these newly introduced vertices and the edges incident to them. Once a newly introduced vertex $r \in R$ is unmatched, the path $b \leadsto r$ is exactly $p^{\uparrow}$.

Modifications to $E_h$ must preserve the constraints of the feasible vertex labeling. For all $l \in T.L$ and $r \in R - T.R$, we have $l.h + r.h > w(l,r)$ when $T$ is unexpandable. The smallest difference is calculated as $\delta = \min\{l.h + r.h - w(l,r) : l \in T.L\ \text{and}\ r \in R - T.R\}$. The new vertex labeling $h'$ is then defined as follows:

\begin{equation}
    v.h' =  
    \begin{cases} 
        v.h - \delta & \text{if} \ v \in T.L \\
        v.h + \delta & \text{if} \ v \in T.R \\
        v.h & \text{otherwise} \\
    \end{cases}
\end{equation}

Next, for all $l \in L$ and $r \in R$, the constraint $l.h' + r.h' \geq w(l, r)$ under $h'$ still holds, all red edges belong to $E_{h'}$, and $T$ is expandable. Because:
\begin{itemize}
    \item For all $l \in T.L, r \in T.R$, we have $l.h' + r.h' = l.h + r.h$, implying that $E_h \cap (T.L \times T.R) \subset E_{h'}$.
    \item For all $l \in L - T.L, r \in R - T.R$, we have $l.h' + r.h' = l.h + r.h$, implying that $E_h \cap ((L - T.L) \times (R - T.R)) \subset E_{h'}$.
    \item All red edges, which belong to $(T.L \times T.R) \cup ((L - T.L) \times (R - T.R))$, are also contained in both $E_h$ and $E_{h'}$. They are absent from the following two items.
    \item For all $l \in T.L, r \in R - T.R$, we have $l.h' + r.h' < l.h + r.h$, but still satisfying $l.h' + r.h' \geq w(l, r)$. Since $E_h \cap (T.L \times (R - T.R)) = \emptyset$, $T$ was previously unexpandable. Now, it can expand along the newly introduced black edges, given by $\{(l, r) : l \in T.L, r \in R - T.R, l.h' + r.h' - w(l,r)= 0\} \subset E_{h'}$.
    \item For all $l \in L - T.L, r \in T.R$, we have $l.h' + r.h' > l.h + r.h$, and naturally $l.h' + r.h' > w(l, r)$. Consequently, black edges in $E_h \cap ((L - T.L) \times T.R)$ disappear in $E_{h'}$.
\end{itemize}

It takes $O(LR) = O(R^2)$ time to calculate $\delta$ with $\min\{l.h + r.h - w(l,r) : l \in T.L\ \text{and}\ r \in R - T.R\}$. However, this can be optimized by converting it to $\delta = \min\{r.slack : r \in R - T.R\}$. Initially, for each $r \in R$, $r.slack$ is set to $\infty$ before the current BFS begins. As $T$ expands, for each $r \in R - T.R$, $r.slack$ is continuously minimized and maintained at $\min\{l.h + r.h - w(l, r) : l \in T.L\}$. Consequently, the overall time for calculating $\delta$ decreases to $O(R)$. Readers might find this step unclear, but specific code implementations can help clarify the process.

Then, the expansion of $T$ continues along the newly introduced edges. Once $p^{\uparrow}$ is identified, the matching can be extended. If $p^{\uparrow}$ has not been found when $T$ is temporarily unexpandable again, the feasible vertex labeling will be adjusted based on $h'$.

After locating $p^{\uparrow}$ and extending the matching, a new BFS is performed from the next vertex in $L$. 

Note that in the code implementation, we update $v.h$ using the operations $v.h \gets v.h + \delta$ or $v.h \gets v.h - \delta$. Therefore, in the subsequent analysis, we treat $v.h$ as a \textbf{variable}.  

\textbf{Lemma 3.1.1}. Throughout Procedure \ref{sec:procedure}, no vertex transitions from matched to unmatched, ensuring that the size of the matching never decreases.

\textit{Proof}. During each BFS:  
\begin{itemize}  
    \item Each adjustment to $h$ ensures that all red edges in $E_h$ remain in the updated set $E_h'$, thereby matched vertices stay matched.
    \item When $p^\uparrow$ is applied, matched vertices may change partners, but they remain matched, as $p^\uparrow$ merely redistributes edges within the matching.  
\end{itemize}  

Consequently, no vertex transitions from matched to unmatched throughout the procedure.

\textbf{Theorem 3.1.2}. We can get an MWPM with Procedure~\ref{sec:procedure}.

\textit{Proof}. For each $l \in L$, if it is unmatched after the initial greedy matching, a BFS starting from it will be performed. Then a $p^\uparrow$ must be found and applied to augment the matching. By Lemma 3.1.1, no vertex transitions from matched to unmatched. Through this iterative process, a perfect matching is gradually achieved, without violating the constraints of the equality graph. As demonstrated in Theorem 2.3.1, this perfect matching is also the MWPM.

\textbf{Corollary 3.1.3}. The initial greedy matching is unnecessary Procedure~\ref{sec:procedure}. 

\textit{Proof}. For each unmatched vertex $l \in L$, a BFS will start with it and augment the matching when the BFS ends.

\textbf{Corollary 3.1.4}.  
Throughout the procedure, for any unmatched vertex $r \in R$, we have $r.h = 0$.  

\textit{Proof}. By Lemma 3.1 and Procedure \ref{sec:procedure}, any vertex $\in R$ is matched as long as it has been introduced into $T$. Thus each unmatched $r \in R$ has not been appeared into $T$, and has no opportunity to increase $r.h$, which means $r.h$ remains 0 from the beginning.

\textbf{Corollary 3.1.5}.
Throughout the procedure, for any vertex $r \in R$, we have $r.h \geq 0$.  

\textit{Proof}. As previously established, the following conditions hold:  
\begin{itemize}
    \item If $(l, r) \notin E_h$, then $l.h + r.h - w(l, r) > 0$.
    \item $\delta = \min\{l.h + r.h - w(l, r) : l \in T.L,\ r \in R - T.R\}$.
\end{itemize}

Thus $\delta > 0$ before we get an MWPM. Then each time we update $r.h$ for every $r \in T.R$ by adding $\delta$, its value strictly increases, thereby confirming the claim with Corollary 3.1.4.  

\newpage
\subsection{Code}

\begin{algorithm}[H]
\caption{Hungarian}
\KwData{Complete bipartite graph $G = (L, R, w)$ requiring $|L| = |R|$}
\KwResult{Maximum weight perfect matching }
\ForEach{$l \in L$}{
    $l.pair \gets \text{nil}$\;
    
    $l.h \gets \max \{ w(l, r) : r \in R \}$\;
}
\ForEach{$r \in R$}{
    $r.pair \gets \text{nil}$\;
    
    $r.h \gets 0$\;
    
    $r.\pi \gets \text{nil}$ \tcp{\color{gray}\ttfamily Parent node of $r$ in the breadth-first tree}
}
\BlankLine
\tcp{\color{gray}\ttfamily Initial greedy matching}
\ForEach{$l \in L$}{
    \ForEach{$r \in R$}{
        \If{$l.h + r.h = w(l, r)$ \textbf{and} $r.pair = \text{nil}$}{
            $l.pair \gets r$\;
            
            $r.pair \gets l$\;
            
            \textbf{break}\;
        }
    }
}
\BlankLine
\ForEach{$l \in L$}{
    \If{$l.pair \neq \text{nil}$}{
        \textbf{continue}\;
    }
    $Q \gets \{ l \}$\;
    
    \ForEach{$l' \in L$}{
        $l'.visited \gets \text{false}$\;
    }
    $l.visited \gets \text{true}$\;
    
    \ForEach{$r \in R$}{
        $r.slack \gets \infty$\; 
        
        $r.visited \gets \text{false}$\;
    }
    \textsc{BFS}($Q$)\;
}

\BlankLine
\Return $\{ (l, r) : l \in L \ \text{and}\ r = l.pair \}$
\end{algorithm}

\label{firstBFS}
\begingroup
\let\origthealgocf\thealgocf
\let\origalgorithmcfname\algorithmcfname
\newcommand{\restorealgocounter}{
    \addtocounter{algocf}{-2}
}

{
    \renewcommand{\thealgocf}{}%
    \renewcommand{\algorithmcfname}{}%
    \SetAlgoCaptionSeparator{}
\newpage
\begin{algorithm}[H] 
\caption*{BFS}
\setcounter{AlgoLine}{0}
\KwIn{Queue $Q$}
\While{\textbf{true}}{
    \Repeat{ $Q$ is empty}{
        $l \gets \text{Dequeue}(Q)$\;
        
        \ForEach{$r \in R$}{
            \If{$r.visited$}{
                \textbf{continue}\;
            }
            
            $d \gets l.h + r.h - w(l, r)$\;
            
            \If{$d = 0$}{
            
                $r.\pi \gets l$\;
                
                \If{\textsc{Advance}($Q, r$)}{
                
                    \Return\;
                    
                }
            }
            
            \ElseIf{$r.slack > d$}{
            
                $r.slack \gets d$\;
                
                $r.\pi \gets l$\;
                
            }
        }
    }
    
    \vspace{10pt}
    $T.L \gets \{ l : l \in L \ \text{and}\ l.visited \}$\;
    
    $T.R \gets \{ r : r \in R \ \text{and}\ r.visited \}$\;
    
    $\delta \gets \min \{ r.slack : r \in R - T.R \}$\;

    \vspace{10pt}
    \ForEach{$l \in T.L$}{
    
        $l.h \gets l.h - \delta$\;
    }
    
    \ForEach{$r \in T.R$}{
    
        $r.h \gets r.h + \delta$\;
        
    }
    
    \ForEach{$r \in R - T.R$}{
    
        $r.slack \gets r.slack - \delta$\;
        
    }
    
    \ForEach{$r \in R - T.R$}{
    
        \If{$r.slack = 0$ \textbf{and} \textsc{Advance}($Q, r$)}{ 
        
            \Return\;
        }
    }
}
\end{algorithm}

\begin{algorithm}[H]
\caption*{Advance}
\KwIn{Queue Q, Vertex $r$}
\KwResult{True if $p^\uparrow$ has been found and applied; False if $p^\uparrow$ has not been found from the root, and the search should continue.}
\setcounter{AlgoLine}{0}
$r.visited \gets \text{true}$\;

$l \gets r.pair$\;

\If{$l \neq \text{nil}$}{

    \text{Enqueue}($Q, l$)\;
    
    $l.visited \gets \text{true}$\;
    
    \Return \textbf{false}\;
}

\tcp{\color{gray}\ttfamily $l = r.pair = nil$ tells $p^\uparrow$ is found, and being applied now.}

\Repeat{$r = $ nil}{

    $l \gets r.\pi$
    
    $prevR \gets l.pair$ \tcp{\color{gray}\ttfamily $l.pair$ is also its parent node on $p^\uparrow$.}
    
    $l.pair \gets r$
    
    $r.pair \gets l$
    
    $r \gets prevR$
}

\Return \textbf{true}\;
\end{algorithm}
}

\restorealgocounter
\renewcommand{\thealgocf}{\origthealgocf}            
\renewcommand{\algorithmcfname}{\origalgorithmcfname}
\endgroup

\subsection{Allowing $|L| < |R|$} \label{sec:allowing}
As mentioned in the introduction~\ref{sec:Introduction}, when seeking a maximum weight matching, we can convert MWM to MWPM by disregarding negatively weighted edges and introducing virtual vertices and edges with zero weight to ensure $|L| = |R|$ and $|E| = |L||R|$. However, this section demonstrates the theorem below.

\textbf{Theorem 3.3.1}. In the non-line-covering variant of the Hungarian algorithm for the MWM problem,  additional virtual vertices are redundant, meaning we only need to ensure $|E| = |L||R|$.

\textit{Proof}. Define $M^{\downarrow}$ as the final imperfect matching got by Procedure \ref{sec:procedure} on a complete weighted bipartite graph with $|L| < |R|$. We still have $|M^{\downarrow}| = |L|$.

We partition the final set $R$ into two subsets:
\begin{itemize}
    \item $R_1$ consists of all matched vertices. Thus, $|R_1| = |M^{\downarrow}| = |L|$.
    \item $R_2$ consists of all unmatched vertices. Thus, $|R_2| = |R| - |L|$.
\end{itemize}

As mentioned in Section "Equality Subgraph"~\ref{sec:EqualitySubGraph}, the weight of any matching $M$ in $G$ can be expressed as:
\begin{equation}
    w(M) = \sum\limits_{(l,r)\in M}w(l,r) \leq \sum\limits_{(l,r)\in M}(l.h + r.h).
\end{equation}

Since each vertex in $L$ must be matched, and by Corollary 3.1.5, we have $r.h \geq 0$ for all $r \in R$, the equation can be simplified to:

\begin{equation}
w(M) \leq  \sum\limits_{l \in L} l.h + \sum\limits_{r \in R} r.h = \sum\limits_{l \in L} l.h + \sum\limits_{r \in R_1} r.h + \sum\limits_{r \in R_2} r.h.
\end{equation}

By Corollary 3.1.4, $\forall r \in R_2,\ r.h = 0$. Substituting this into the new equation simplifies it to:
\begin{equation}
    w(M) \leq \sum\limits_{l \in L} l.h + \sum\limits_{r \in R_1} r.h = w(M^{\downarrow}).
\end{equation}

Thus, $M^{\downarrow}$ is an MWM in the complete bipartite graph where $|L| < |R|$ and each vertex in $L$ must be matched.

\subsection{Variance of Time Complexity} \label{sec:variance}
For each vertex $l \in L$, if $l$ remains unmatched after the initial greedy matching, we have the following:  
\begin{itemize}  
    \item The BFS process starting from $l$ runs in $O(E)$ time.  
    \item The time complexity for $h$ adjustments is $O(LR)$ without additional vertices and $O(R^2)$ with additional vertices, proved by follows:
    \begin{itemize}
        \item Without additional virtual vertices, the vertex labeling $h$ is adjusted $O(L)$ times. Since at most $|L|$ vertices in $R$ are matched, and as soon as an unmatched vertex in $R - T.R$ is introduced into $T$, $p^\uparrow$ is formed.
        \item When virtual vertices are added, the number of $h$ adjustments is $O(R)$. Although it is actually $O(L)$, we do not analyze this further, as it is either a lower-order or the same-order term, as explained below.
        \item Each adjustment requires $\Theta(R)$ time.
    \end{itemize}
\end{itemize}  

As a result, regardless of whether the initial greedy matching is introduced, we conclude the following;
\begin{itemize}  
    \item When virtual vertices and their incident edges are added, $|E|$ increases from $|L||R|$ to $|R|^2$. Each virtual vertex $l$ is either matched in the initial greedy matching, or matched within $O(R)$ time during BFS as it can introduce all unmatched vertices $r \in R$ at the start because $l.h + r.h - w(l, r) = 0$. Consequently, the total time complexity is $O(L(E + R^2) + (R - L)R) = O(LR^2)$.  
    \item When no virtual vertices are added, the total time complexity is $O(L(E+ LR)) = O(L^2R)$.  
\end{itemize}

\section{Extensional Hungarian Algorithm for The MWM Problem}

\subsection{Introduction}
Previously, we demonstrated that the traditional non-line-covering variant of the Hungarian algorithm can be applied to a complete weighted bipartite graph $G = (L, R, E, w)$ to obtain an MWM, requiring only additional edges to ensure $|E| = |L||R|$. As is known, edges with non-positive weights can be ignored in an MWM problem. In this section, we introduce the following new optimizations:  
\begin{itemize}  
    \item Restrict $|E|$ to at most $|L|^2$.  
    \item Remove additional virtual edges, allowing $|E| < |L||R|$.  
    \item Eliminate consecutive $h$ adjustments.  
\end{itemize}  

\subsection{Removing Redundant Edges}

\textbf{Theorem 4.2.1}. Let an MWM in the original weighted bipartite graph be $M_1$. For each $l \in L$, if the size of $G.Adj[l]$ exceeds $|L|$, we retain only the top $|L|$ heaviest edges adjacent to $l$. Let an MWM in the new graph be $M_2$. We have $w(M_1) = w(M_2)$.

\textit{Proof}. For any pair $(l, r) \in M_1$, suppose there exist at least $|L|$ edges incident to $l$ with weights \textbf{unstrictly} greater than $w(l, r)$. Since $|M_1| \leq |L|$, at least one edge $(l, r')$ satisfies the following:
\begin{itemize}
    \item $(l, r') \in E$
    \item $(l, r') \notin M_1$
    \item $w(l, r') \geq w(l, r)$
    \item $r'$ is unmatched
\end{itemize}

By replacing $(l, r)$ with $(l, r')$, we obtain a matching $M'$ with $w(M') \geq w(M_1)$. $w(M') > w(M_1)$ contradicts the weight maximality of $M_1$, and $w(M') = w(M_1)$ tells $(l, r)$ is replaceable. Therefore, when computing an MWM, we can safely discard all but the top $|L|$ heaviest edges adjacent to each $l$ without affecting the maximality of the result.

\textbf{Corollary 4.2.2}. For the MWM problem, the number of edges in a bipartite graph can be limited to at most $|L|^2$.  

\textit{Proof}. This is apparently confirmed by Theorem 4.2.1.

\subsection{Removing Additional Virtual Edges} \label{sec:ReplaceAddedEdges}

\textbf{Lemma 4.3.1}. For each $l \in L$ and $r_1, r_2 \in R$, if $r_1$ is unmatched, $r_1 \neq r_2$, and $l.h + r_1.h - w(l,r_1) \leq l.h + r_2.h - w(l, r_2)$, then the edge $(l, r_1)$ can be prioritized while $(l, r_2)$ is ignorable when $l$ appears in $T$.

\textit{Proof}. When $l$ appears in $T$, we analyze $r_1$ and $r_2$ as follows:
\begin{itemize}
    \item If $l.h + r_1.h - w(l,r_1) = 0$, we can introduce $r_1$ into $T$, thereby forming $p^\uparrow$ without considering about $(l, r_2)$. 
    \item If $l.h + r_1.h - w(l, r_1) > 0$, their respective slack values are updated as follows:
    \begin{itemize}
        \item $r_1.slack' = \min(r_1.slack,\ l.h + r_1.h - w(l, r_1))$ 
        \item 
        \begin{equation}
            r_2.slack' =  
                \begin{cases} 
                    \min(r_2.slack,\ l.h + r_2.h - w(l, r_2)) & \text{if}\ (l, r_2)\ \text{is used} \\
                    r_2.slack & \text{if}\ (l, r_2)\ \text{is ignored}
                \end{cases}
        \end{equation}
    \end{itemize}
    
    If $r_1.slack \leq r_2.slack$, then using $(l, r_2)$ leads to $r_1.slack' \leq r_2.slack'$, and ignoring $(l, r_2)$ leads to $r_1.slack' \leq r_1.slack \leq r_2.slack = r_2.slack'$. This means we always get $r_1.slack' \leq r_2.slack'$, and ignoring $(l, r_2)$ does not affect the priority of introducing $r_1$, nor does it impact $\delta$ in the event of an $h$ adjustment. Even if $r_2.slack$ is not minimized, $p^\uparrow$ is still formed when $r_1$ enters $T$ first and the BFS ends, making $r_2$ redundant until its $slack$ becomes smaller than that of $r_1$.
    
    If $r_1.slack > r_2.slack$, then:
    \begin{itemize}
        \item If $r_2.slack \leq l.h + r_2.h - w(l, r_2)$, ignoring $(l, r_2)$ makes no difference.
        \item If $r_2.slack > l.h + r_2.h - w(l, r_2)$, then using $(l, r_2)$ leads to $r_1.slack' \leq l.h +r_1.h - w(l, r_1) \leq l.h +r_2.h - w(l, r_2) = r_2.slack'$, and ignoring $(l, r_2)$ leads to $r_1.slack' \leq l.h + r_1.h - w(l, r_1) \leq l.h + r_1.h - w(l, r_2) < r_2.slack = r_2.slack'$. Similarly, we always have $r_1.slack' \leq r_2.slack'$, allowing us to ignore $(l, r_2)$.
    \end{itemize}
\end{itemize}

Overall, we can ignore each edge $(l, r_2)$ with the given conditions in Lemma 4.3.1.

\textbf{Theorem 4.3.2}. Let $r'$ be the first unmatched vertex in $R$. Suppose that for each $l \in L$, $w(l, r') = 0$ if $(l, r') \notin E$. Then, Procedure~\ref{sec:procedure} can be executed correctly on a general incomplete weighted bipartite graph.

\textit{Proof}. In Procedure \ref{sec:procedure}, for each virtual edge $(l, r)$, we observe the following properties:
\begin{itemize}
    \item By Corollary 3.1.4, $r'.h = 0$. 
    \item Since $(l, r')$ may not be virtual, we have $w(l, r') \geq 0$ instead of $w(l, r') = 0$.
    \item By Corollary 3.1.5, $r.h \geq 0$. 
    \item $w(l, r) = 0$.
\end{itemize}

Thus, we conclude that $l.h + r'.h - w(l, r') \leq l.h + r.h - w(l, r)$ since this simplifies to $-w(l,r') \leq 0 \leq r.h$. By Lemma 4.3.1 and under the assumption that for each $l \in L$, $w(l, r') = 0$ if $(l, r') \notin E$, any vertex $r \in R$ can be ignored if it is neither $r'$ nor adjacent to $l$. 

Consequently, an edge count less than $|L||R|$ is permissible. Moreover, by Theorem 3.3.1, the condition $|L| < |R|$ is also allowed. This ensures that Procedure~\ref{sec:procedure} can be correctly executed on a general incomplete weighted bipartite graph.

\subsection{Eliminating Consecutive $h$ Adjustments} \label{sec:replaceCon}
\textbf{Theorem 4.4.1}. In each BFS, consecutive $h$ adjustments can be done at once with code in Section \ref{Code2}.

\textit{Proof}. In each BFS before $p^\uparrow$ is found, we define additional properties as follows:  

\begin{itemize}  
    \item Let $i$ denote the stage number starting from $0$. Each time we calculate $\delta$, let $i$ increase by $1$.
    \item For each $l \in T.L$ and $r \in T.R$, let $l.i$ and $r.i$ denote the stage number at which $l$ is introduced into $T$. We have $l.i, r.i \in [0, |L|)$.  
    \item Let $\Delta$ store each $\delta$ in chronological order.  
    \item Let $\sum\delta$ be the cumulative sum of all $\delta$.  
    \item For each vertex $r \in R - T.R$, we replace $r.slack$ with $r.slack' = r.slack + \sum\delta$. This ensures that the uniform decrement of $r.slack$ in the previous BFS algorithm~\ref{firstBFS} can be efficiently handled by incrementing $\sum\delta$.  
    \item Let $r^\star$ be the \textbf{unmatched} vertex in $R - T.R$ with the smallest $slack'$.  
    \item Let $F$ be the Fibonacci heap containing \textbf{matched} vertices $r \in R - T.R$, sorted by $r.slack'$.  
\end{itemize}  

Since $\delta = \min\{r.slack : r \in R - T.R\}$ in Procedure~\ref{sec:procedure}, we now define $\delta$ as follows:  

\begin{equation}  
    \delta =  
    \begin{cases}  
        r^\star.slack' - \sum\delta & \text{if } F \text{ is empty} \\  
        \min(r^\star.slack',\ F.min.slack') - \sum\delta & \text{otherwise}  
    \end{cases}  
\end{equation}  

We do not adjust $h$ immediately after calculating $\delta$. Specifically:  

\begin{enumerate}  
    \item Insert $\delta$ to $\Delta$ and update $\sum\delta$.  
    \item If $r^\star.slack' - \sum\delta = 0$, then $r^\star.slack = 0$ and $p^\uparrow$ is found.  
    \item If $r^\star.slack' - \sum\delta > 0$, it implies that no unmatched vertices in $R - T.R$ can be introduced into $T$ at the time. We extract the minimum elements $r$ from $F$ and continue the BFS, as they have the smallest $slack'$ and must satisfy $r.slack' - \sum\delta = 0$.
\end{enumerate}  

The search continues until $p^\uparrow$ is found. We adjust $h$ only after applying $p^\uparrow$. Vertices with smaller $l.i$ or $r.i$ receive a larger portion of the $\delta \in \Delta$. This adjustment can be efficiently performed by computing the suffix sums of $\Delta$. Further details are provided in the sub-algorithm Advance.  

\newpage

\subsection{Code} \label{Code2}
\begin{algorithm}[H]
\caption{Kwok}
\setcounter{AlgoLine}{0}
\KwData{Bipartite graph $G = (L, R, E, w)$ with adjacency list. }
\KwResult{Maximum weight matching}
\BlankLine

Remove non-positively weighted edges $\in E$

\ForEach{$l \in L$}{

    $l.pair \gets \text{nil}$\;
    
    $l.h \gets \max \{ w(l,r) : (l, r) \in E \}$\;
    
    \If{\textnormal{the size of} $G.Adj[l] > |L|$}{
    
        Retain only the top $|L|$ heaviest edges\ incident to $l$ with the $i$th order statistic algorithm \cite{Cormen2022}
    }
}

\ForEach{$r \in R$}{

    $r.pair \gets \text{nil}$\;
    
    $r.h \gets 0$\;
    
    $r.\pi \gets \text{nil}$\;
}

\BlankLine 
\tcp{\color{gray}\ttfamily Initial greedy matching}
\ForEach{$l \in L$}{

    \ForEach{$r \in G.Adj[l]$}{
    
        \If{$l.h + r.h = w(l, r)$ \textbf{and} $r.pair = \text{nil}$}{
        
            $l.pair \gets r$\;
            
            $r.pair \gets l$\;
            
            \textbf{break}\;
        }
    }
}

\BlankLine

\ForEach{$l \in L$}{

    \If{l.pair $\neq$ nil}{
    
        \textbf{continue};
    }
    
    \BlankLine
    \ForEach{$l \in L$}{
    
        $l.i \gets \text{nil}$\;
    }
    
    \ForEach{$r \in R$}{
    
        $r.i \gets \text{nil}$\;
        
        $r.slack' \gets \infty$\;
    }
    
    \BlankLine
    $l.i \gets 0$\;
    
    $Q \gets \{ l \}$\;
    
    \textsc{BFS}($Q$)\;
}

\Return $\{(l, r) : l \in L \ \text{and}\ r = l.pair \ \text{and}\ (l, r) \in E\}$ \tcp{\color{gray}\ttfamily Ensure $(l, r) \in E$ to remove virtual pairs. }
\end{algorithm}

\newpage  
\vspace*{-1cm}
\renewcommand{\thefootnote}{}
\footnote{$^1$In this part, each time we use $l.h$ or $r.h$, $l$ is newly introduced into $T$ while $r$ remains outside $T$. Therefore, they do not correspond to any $\delta \in \Delta$, and the conditions among $l.h$, $r.h$ and $w$ remain unaffected by the deferred update for $h$.}
\footnote{$^2$There is not enough space here, so we place this footnote at the top of the next page.}
\begin{algorithm}[H]  
\caption*{BFS}
\setcounter{AlgoLine}{0}
\footnotesize
\KwIn{Queue $Q$}

$r' \gets \text{the first unmatched vertex in}\ R$\;

$F \gets$ an empty Fibonacci heap for containing vertices, sorted by their $slack'$\;

$\sum\delta \gets 0$\;

$\Delta \gets$ an empty list\;

$i \gets 0$\;

$r^\star \gets r'$\;

$^1$
\While{\textbf{true}}{

    \Repeat{$Q$ is empty}{
    
        $l \gets \text{Dequeue}(Q)$\; 
          
        $^2$
        \If{$l.h = 0$}{
        
            $r'.\pi \gets l$\;
            
            \textsc{Advance}($Q, F, r', \Delta, i$)\;
            
            \Return\;
        }

        \If{$r'.slack' > l.h + \sum\delta$}{
        
            $r'.slack' \gets l.h + \sum\delta$\;
            
            $r'.\pi \gets l$\;
        }

        \If{$r^\star.slack' >  r'.slack'$}{
        
            $r^\star \gets r'$\;
        }
        
        \BlankLine
        \ForEach{$r \in G.Adj[l]$}{
        
            \If{$r'.i \neq \text{nil}$}{ \tcp{\color{gray}\ttfamily This tells $r' \in T.R$, thus we skip.}
            
                \textbf{continue}\;
            }
            
            $d \gets l.h + r.h - w(l,r)$\;

            \If{$d = 0$}{
            
                $r.\pi \gets l$\;
                
                \If{\textsc{Advance}($Q, F, r, \Delta, i$)}{
                
                    \Return\;
                }            
            }
            
            \ElseIf{$r.slack' > d + \sum\delta$}{
            
                $r.slack' \gets d + \sum\delta$\;
                
                $r.\pi \gets l$\;

                \If{$r.pair = \text{nil}$}{
                
                    \If{$r^\star.slack' > r.slack'$}{
                    
                        $r^\star \gets r$\;
                    }
                }
                
                \ElseIf{$r \in F$}{
                
                    Notify $F$ that $r.slack'$ has been decreased\;
                }
                
                \Else{ Insert $r$ into $F$\; }
            }
        }
    }
    
    \If{\textsc{Introduce}($Q, F, r^\star, \Delta, i, \sum\delta$)}{
    
        \Return\;
    }
}
\end{algorithm}

\newpage
\textbf{Supplementary footnote for the previous page:} it's inefficient to get $w(l, r')$ and check whether $(l, r') \in E$ directly without first constructing an adjacency matrix. Instead, we use the following alternative approach:  
\begin{itemize}
    \item If $(l, r') \notin E$, we can suppose $w(l, r') = 0$ according to Section~\ref{sec:ReplaceAddedEdges}. In this case, the equation $l.h + r'.h - w(l, r') = 0$ simplifies to $l.h = 0$, which allows the introduction of $r'$. Then, if $l.h \neq 0$, $r'.slack'$ is correctly minimized with $l.h + r'.h - w(l, r') + \sum\delta = l.h + \sum\delta$.
    
    \item If $(l, r') \in E$, it's correct to skip at the time, and process $(l, r')$ in the subsequent "foreach" loop. However, we have $w(l, r') > 0$ and the equation $l.h + r'.h - w(l, r') = l.h - w(l,r') \geq 0$ implies that $l.h > 0$ and $r'$ won't be processed by the function \textsc{Advance} on the 12th line. Then, $r'.slack$ is incorrectly minimized with $l.h + \sum\delta$ which is larger than $l.h + r'.h - w(l,r') + \sum\delta$. However, $r'$ will be processed in the subsequent "foreach" loop unless $p^\uparrow$ is formed beforehand. In that case, $r'$ is introduced, or $r'.slack$ is correctly minimized with $d + \sum\delta = l.h + r'.h - w(l, r') + \sum\delta$.
\end{itemize}

\begin{algorithm}
\caption*{Introduce}
\setcounter{AlgoLine}{0}
\KwIn{Queue $Q$, Fibonacci Heap $F$, Vertex $r^\star$, List $\Delta$, Int $i, \sum\delta$}
\KwResult{True if the function \textsc{Advance} has been executed and returned True.}

    $i \gets i + 1$\;
    
    $\delta \gets r^\star.slack' - \sum\delta$\;
    
    \BlankLine
    \If{$F$ \textnormal{is not empty}}{
    
        $\delta \gets \min(\delta, F.min.slack' - \sum\delta)$\;
    }
    
    \BlankLine
    Insert $\delta$ to $\Delta$\;
    
    $\sum\delta \gets \sum\delta + \delta$\;

    \BlankLine
    \If{$r^\star.slack' - \sum\delta = 0$}{
    
        \textsc{Advance}($Q,\ F,\ r^\star, \Delta, i$)\;
        
        \Return \textbf{true}\;
    }

    \BlankLine
    \While{$F$ \textnormal{is not empty} \textbf{and} $F.min.slack' - \sum\delta = 0$}{
    
        $r \gets$ ExtractMin($F$)\;
        
        \tcp{\color{gray}\ttfamily $r$ is matched, and thus \textsc{Advance} must return false here.}
        
        \textsc{Advance}($Q,\ F,\ r, \Delta, i$)\; 
    }
    
    \Return \textbf{false}\;
\end{algorithm}

\newpage
\begin{algorithm}[H] 
\caption*{Advance}
\setcounter{AlgoLine}{0}
\KwIn{Queue $Q$, Fibonacci Heap $F$, Vertex $r$, List $\Delta$, Int $i$}
\KwResult{True if $p^\uparrow$ has been found and applied; False if $p^\uparrow$ has not been found from the root, and the search should continue.}

$r.i \gets i$\;

\If{$r \in F$}{

    Delete $r$ from $F$\;
}

\BlankLine
$l \gets r.pair$\;

\If{$l \neq \text{nil}$}{

    Enqueue($Q,\ l$)\;
    
    $l.i \gets i$\;
    
    \Return \textbf{false}\;
}
\BlankLine

\tcp{\color{gray}\ttfamily Augment the matching by applying $p^\uparrow$.}

\Repeat{$r = \text{nil}$}{

    $l \gets r.\pi$\;
    
    $prevR \gets l.pair$\; 
    
    $l.pair \gets r$\;
    
    $r.pair \gets l$\;
    
    $r \gets prevR$\;
}
\BlankLine

\If{$\Delta$ is empty}{
    \Return \textbf{true}\;
}

\tcp{\color{gray}\ttfamily Update $h$ with suffix sums of $\Delta$.}
$S[0:|\Delta|-1] \gets$ an new array cloned from $\Delta$\;

\ForEach{$j = |\Delta|-2$ \textbf{to} $0$}{

    $S[j] \gets S[j] + S[j+1]$\;
}

\ForEach{$l' \in L$}{

    \If{$l'.i \neq$ nil \textbf{and} $l'.i \neq i$}{
    
        $l'.h \gets l'.h - S[l'.i]$\;
    }
}

\ForEach{$r' \in R$}{

    \If{$r'.i \neq$ nil \textbf{and} $r'.i \neq i$}{
    
        $r'.h \gets r'.h + S[r'.i]$\;
    }
}

\BlankLine
\Return \textbf{true}\;
\end{algorithm}

\subsection{Time Complexity} 
The preparation runs in $\Theta(E)$ time for limiting the edge count, setting the initial values of $l.h$ and $r.h$, and executing the initial greedy matching.

For each $l \in L$, we assume $l$ is unmatched after the initial greedy matching, and focus on the process of searching for $p^\uparrow$ starting from $l$ as follows:  
\begin{itemize}  
    \item The initializations of $l.i$, $r.i$, and $r.slack'$ take $\Theta(R)$ time.  
    \item The \textbf{foreach} loop in BFS is iterated $O(\min(L^2, E))$ times.  
    \item $F$ contains matched vertices in $R - T.R$. Specifically, only \textbf{matched} vertices in $R - T.R$ have the opportunity to be inserted into $F$ and may be later extracted into $T$. Since in a Fibonacci heap, the \texttt{Insert} and \texttt{DecreaseKey} operations take $O(1)$ amortized time, and \texttt{ExtractMin} takes $O(\log n)$ amortized time, we conclude that all operations involving $F$ take $O(L \log L)$ time.  
    \item The deferred update for $h$ takes $\Theta(R)$ time.  
    \item The application of $p^\uparrow$ takes $O(2L) = O(L)$ time.  
\end{itemize}  

Therefore, all the above operations take $O(\min(L^2, E) + L \log L) = O(\min(L^2,\ E + L\log L))$ time. Consequently, considering all $p^\uparrow$ starting from vertices in $L$, the total worst-case time complexity of the algorithm is $O(\min(L^3 + E,\ LE + L^2\log L))$.  

\subsection{Further Optional Optimization}
For each $l \in L$:

\begin{enumerate}
    \item Sort $G.Adj[l]$ by weight in descending order.
    \item During BFS, when an unmatched vertex $r \in R$ is encountered:
    \begin{itemize}
        \item If $l.h + r.h - w(l,r) = 0$, $p^\uparrow$ can be formed immediately.
        \item If $l.h + r.h - w(l,r) > 0$, after attempting to update $r.slack'$ and $r.\pi$, the remaining vertices $r' \in G.Adj[l]$ can be ignored. This is justified as follows:
        \begin{itemize}
            \item By Corollary 3.1.4, $r.h = 0$.
            \item By Corollary 3.1.5, $r'.h \geq 0$.
            \item $l.h + r.h - w(l, r) = l.h - w(l, r) > 0$.
            \item $w(l, r') \leq w(l, r)$.
        \end{itemize}
        Consequently, we have $r'.h - w(l,r') \geq -w(l,r)$ which leads to $l.h + r'.h - w(l, r') \geq l.h + r.h - w(l, r)$. By Lemma 4.3.1, all the subsequent vertices $r' \in G.Adj[l]$ can be safely ignored.
    \end{itemize}
\end{enumerate}

This does reduce the count of visited edges in each BFS, but it's challenging to quantify the exact impact on runtime. Experimental results indicate that this optimization provides limited benefits for sparse graphs, while for dense graphs, the sorting process may introduce substantial computational overhead, potentially offsetting the original superior average performance. However, in dynamic scenarios where the adjacency list can be efficiently reordered in descending order by weight within linear time complexity, this optimization becomes a worthwhile consideration.  

\section{Experiments}

\subsection{Actual Implementation}

The implementation code of Kwok's algorithm is available at \url{https://github.com/ShawxingKwok/Kwok-algorithm}. Note that after extensive experiments and comparisons, we choose to only keep the optimization of replacing added virtual edges~\ref{sec:ReplaceAddedEdges}. The worst-case time complexity of this implementation is $O(LE + LR\min(L, \frac{N}{p}))$, where the maximum weight is $N$ and the weights are represented with a precision of $p$. For example, if the weights fall within the range $[1.00, 10.00]$ with a precision of $0.01$, then $p = 0.01$. If there exist irrational fractional weights, then $p = o(1)$, meaning $O(LE + LR\min(L,\frac{N}{p})) = O(L^2R)$. However, the experimental performance is well enough.

\subsection{Run Time Contrast among Algorithms with Different $|E|$ and $|L| : |R|$ }

\textit{Notes}: 
\begin{itemize}
    \item The test environment uses an Apple M1 Pro chip with 16 GB of memory.
    \item $|L|$ is fixed at $1000$.
    \item The time unit for all measurements is milliseconds.
    \item Weights are integers and randomly distributed within the range $[1, |R|]$.
    \item ‘MCMF \& Dijkstra’ is the algorithm introduced in Section~\ref{sec:MCMF}.
    \item The programming language used is uniformly Kotlin-JVM. The source code for some algorithms is selected from popular websites listed below and converted to Kotlin-JVM.  
    \begin{itemize}  
        \item Line-covering: \url{https://cp-algorithms.com/graph/hungarian-algorithm.html}.  
        \item Non-line-covering:\url{https://oi-wiki.org/graph/graph-matching/bigraph-weight-match/}. We have added the implementation of the initial greedy matching. $\downarrow$ implies the absence of additional vertices.
    \end{itemize}  
    \item The number after $\pm$ is the sample standard deviation $\sqrt{\frac{1}{n-1}\sum\limits_{i=1}^n (x_i - \overline{x})^2}$.
    \item In each round, a random bipartite graph is generated under specified conditions and shared among different algorithms. The run time of each algorithm is then recorded. After 10 rounds, the average run time of each algorithm is summarized in tables.
    \item As $|R|$ increases, observed runtime may deviate from the expected time complexity. This deviation is partly attributed to the increasing effectiveness of the initial greedy matching process.
    \end{itemize}

\begin{table*}[!http]
  \caption{Runtime for Different Ratios of $|L| : |R|$ While Keeping $|E| = 0.5 |L| \lg |R|$}
  \centering
  \resizebox{\textwidth}{!}{%
  \begin{tabular}{cccccc}
    \toprule
    |L| : |R| & MCMF \& Dijkstra & line-covering & non-line-covering & non-line-covering $\downarrow$ & Kwok \\
    \midrule   
    1 : 1  & 375.72 $\pm$ 3.10 & 119.89 $\pm$ 16.30 & 14.45 $\pm$ 0.71 & 16.96 $\pm$ 0.82 & 0.89 $\pm$ 0.18 \\
    1 : 2  & 466.78 $\pm$ 1.32 & 11.83 $\pm$ 0.86 & 23.50 $\pm$ 1.82 & 9.68 $\pm$ 0.51 & 0.25 $\pm$ 0.01 \\
    1 : 4  & 568.53 $\pm$ 3.53 & 17.22 $\pm$ 0.95 & 70.79 $\pm$ 1.50 & 12.06 $\pm$ 0.68 & 0.17 $\pm$ 0.01 \\
    1 : 8  & 633.61 $\pm$ 13.35 & 31.48 $\pm$ 0.46 & 286.57 $\pm$ 14.62 & 18.30 $\pm$ 0.88 & 0.16 $\pm$ 0.01 \\
    \bottomrule
  \end{tabular}%
  }
  \vspace{1em}
  \textit{Note}: The graph is sparse.
\end{table*}

\begin{table*}[!http]
  \caption{Runtime for Different Ratios of $|L| : |R|$ While Keeping $|E| = 10 |L| \lg |R|$}
  \centering
  \resizebox{\textwidth}{!}{%
  \begin{tabular}{cccccc}
    \toprule
    |L| : |R| & MCMF \& Dijkstra & line-covering & non-line-covering & non-line-covering $\downarrow$ & Kwok \\
    \midrule   
    1 : 1  & 1228.66 $\pm$ 39.66 & 77.68 $\pm$ 8.35 & 78.46 $\pm$ 7.92 & 78.38 $\pm$ 7.25 & 20.82 $\pm$ 1.41 \\
    1 : 2  & 1786.28 $\pm$ 12.95 & 12.26 $\pm$ 1.55 & 21.58 $\pm$ 2.28 & 10.42 $\pm$ 0.45 & 2.49 $\pm$ 0.14 \\
    1 : 4  & 3027.57 $\pm$ 28.91 & 18.79 $\pm$ 1.97 & 60.14 $\pm$ 11.12 & 12.58 $\pm$ 0.70 & 2.37 $\pm$ 1.87 \\
    1 : 8  & 5189.36 $\pm$ 86.40 & 39.99 $\pm$ 17.44 & 209.75 $\pm$ 50.45 & 20.06 $\pm$ 1.14 & 1.53 $\pm$ 0.19 \\
    \bottomrule
  \end{tabular}%
  }
  \vspace{1em}
  \textit{Note}: The graph is dense.
\end{table*}

\begin{table*}[!http]
  \caption{Runtime for Different Ratios of $|L| : |R|$ While Keeping $|E| = \frac{|L||R|}{10}$}
  \centering
  \resizebox{\textwidth}{!}{%
  \begin{tabular}{cccccc}
    \toprule
    |L| : |R| & MCMF \& Dijkstra & line-covering & non-line-covering & non-line-covering $\downarrow$ & Kwok \\
    \midrule   
    1 : 1  & 1314.35 $\pm$ 77.10 & 85.07 $\pm$ 14.91 & 81.00 $\pm$ 8.01 & 80.29 $\pm$ 6.71 & 21.99 $\pm$ 1.69 \\
    1 : 2  & 2816.16 $\pm$ 99.92 & 13.33 $\pm$ 1.04 & 27.79 $\pm$ 12.19 & 11.08 $\pm$ 0.94 & 4.64 $\pm$ 0.21 \\
    1 : 4  & 6218.67 $\pm$ 123.12 & 28.08 $\pm$ 12.48 & 67.81 $\pm$ 28.90 & 16.55 $\pm$ 2.67 & 5.46 $\pm$ 0.38 \\
    1 : 8  & 14527.20 $\pm$ 338.62 & 53.62 $\pm$ 12.69 & 220.33 $\pm$ 74.02 & 25.41 $\pm$ 6.92 & 8.91 $\pm$ 0.89 \\
    \bottomrule
  \end{tabular}%
  }
  \vspace{1em}
  \textit{Note}: The graph is dense.
\end{table*}

\begin{table*}[!http]
  \caption{Runtime for Different Ratios of $|L| : |R|$ While Keeping $|E| = \frac{|L||R|}{2}$}
  \centering
  \resizebox{\textwidth}{!}{%
  \begin{tabular}{cccccc}
    \toprule
    |L| : |R| & MCMF \& Dijkstra & line-covering & non-line-covering & non-line-covering $\downarrow$ & Kwok \\
    \midrule   
    1 : 1  & 6231.45 $\pm$ 335.45 & 95.36 $\pm$ 38.15 & 75.38 $\pm$ 34.43 & 74.83 $\pm$ 35.59 & 53.88 $\pm$ 6.08 \\
    1 : 2  & 10522.30 $\pm$ 1167.23 & 28.22 $\pm$ 9.23 & 30.85 $\pm$ 6.29 & 17.52 $\pm$ 4.35 & 17.35 $\pm$ 1.13 \\
    1 : 4  & 20100.31 $\pm$ 573.18 & 21.47 $\pm$ 0.65 & 54.40 $\pm$ 4.71 & 14.64 $\pm$ 0.79 & 35.83 $\pm$ 35.34 \\
    1 : 8  & 39733.11 $\pm$ 2668.03 & 37.53 $\pm$ 0.72 & 501.89 $\pm$ 284.55 & 24.58 $\pm$ 1.59 & 73.27 $\pm$ 51.08 \\
    \bottomrule
  \end{tabular}%
  }
  \vspace{1em}
  \textit{Note}: The graph is very dense.
\end{table*}

\newpage
\subsection{Expectation Estimate} \label{sec:expect}
\textit{Notes:}
\begin{itemize}
    \item $|L| = |R|$.
    \item Weights are integers and randomly distributed within the range $[1, |R|^2]$.
    \item Each number represents the average value computed from 10 different random graphs with the same key parameters.
\end{itemize}

\begin{figure}
    \centering
    \includegraphics[width=.9\linewidth]{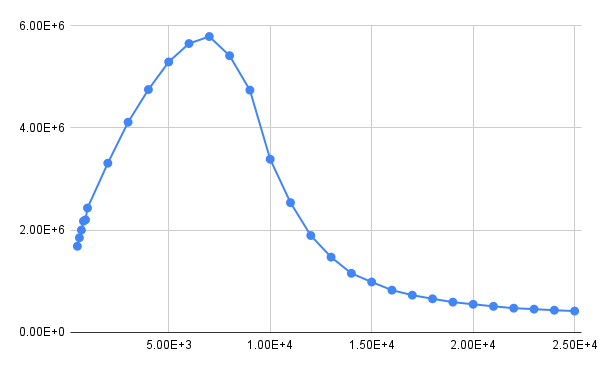}
    \caption{Total number of visited edges when $|E| = 100,000$ and $|L|$ varies from 500 to 1000 in steps of 100, and from 1000 to 25,000 in steps of 1,000}
    \label{fig:fixedE}
\end{figure}

\begin{figure}
    \centering
    \includegraphics[width=.9\linewidth]{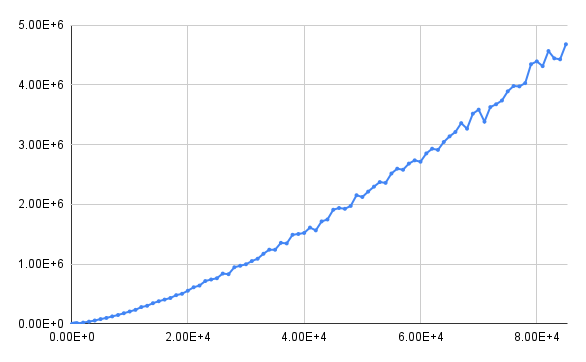}
    \caption{Maximum total number of visited edges when $|E|$ varies from 100 to 1000 in steps of 100, and from 1000 to 85,000 in steps of 1,000.}
    \label{fig:variableE}
\end{figure}

\newpage

To analyze the expected performance of our proposed algorithm, we focus on the fact that each BFS terminates as soon as $p^{\uparrow}$ is formed. However, it's challenging to establish a precise upper bound on the expectation, even on a random graph, as certain algorithmic operations can disrupt the randomness.

In our experiments, even if we remove the optimization in Section~\ref{sec:replaceCon}, the number of total $h$ adjustments remains in $[\frac{|L|}{2}, 3|L|]$. Thus, we focus on the total number of visited edges, as shown in Fig \ref{fig:fixedE} and Fig \ref{fig:variableE}. Figure 9 reveals that there is an interesting maximum number when $|L|$ varies while keeping $|E|$ fixed. In the experiment generating Figure 10, for each fixed $|E|$, we treat $|L|$ as a variable and compute the maximum total number of visited edges.

These statistics suggest that we can establish an upper bound of $O(E^{1.4})$ for the total number of visited edges. Overall, we estimate that our proposed algorithm runs in an average time of $O(E^{1.4} + LR)$ on random weighted bipartite graphs.

\section{Conclusion}
This paper focuses on the MWM problem on a weighted bipartite graph $G = (L, R, E, w)$. We assume $|L| \leq |R|$ and achieve the following advancements:  
\begin{itemize}  
    \item When applying the non-line-covering variant of the Hungarian algorithm on a general weighted bipartite graph, it's proved that adding virtual vertices is unnecessary. This leads to an improved time complexity of $O(L^2R)$, compared to the previous $O(LR^2)$.  
    \item It introduces a novel extension of the non-line-covering variant that runs in $O(\min(L^3+E,\ LE + L^2\log L))$ time, improving upon the previously best-known bound of $O(LE + LV\log V)$.  
    \item The suggested implementation code is simplified and publicly available at \url{https://github.com/ShawxingKwok/Kwok-algorithm}. Although it has a higher theoretical worst-case time complexity of $O(LE + LR\min(L, \frac{N}{p}))$, experimental results indicate that its average running time on random graphs can be estimated as $O(E^{1.4} + LR)$.
\end{itemize}  

These advancements make the algorithm particularly well-suited for applications such as resource allocation, scheduling, and optimization tasks in industrial and networked systems.  

\section*{Acknowledgments}
The author thanks Yiyang Li and another anonymous reader for their invaluable feedback and constructive suggestions.

\bibliographystyle{unsrt}  
\bibliography{sample-manuscript}

\end{document}